\documentclass{revtex4}

\usepackage[small]{caption}
\usepackage{amsmath,amsfonts,amscd,amssymb,epsf, epsfig,mathrsfs}\usepackage{graphicx}

\newcommand{\pa}{\partial}

\begin{document}
\title{Finite temperature Fermionic Casimir interaction in Anti-de Sitter spacetime}
\author{L.P. Teo}\email{ LeePeng.Teo@nottingham.edu.my}\address{Department of Applied Mathematics, Faculty of Engineering, University of Nottingham Malaysia Campus, Jalan Broga, 43500, Semenyih, Selangor Darul Ehsan, Malaysia. }

\begin{abstract}
 We study the finite temperature Casimir interactions on two parallel boundaries in the anti-de Sitter spacetime AdS$_{D+1}$ induced by the vacuum fluctuations of a massive fermionic field  with MIT bag boundary conditions. As in the Minkowski spacetime, the Casimir interaction always  tends to attract the two boundaries to each other at any temperature and for any mass. For bosonic fields, it is well known that the high temperature leading term of the Casimir interaction is linear in temperature. However, for fermionic fields,  the Casimir interaction decays exponentially at high temperature due to the absence of  zero Matsubara frequency.
\end{abstract}

\keywords{Casimir interaction, finite temperature, anti-de Sitter spacetime, massive fermionic field}
 \maketitle

The studies of Casimir effect in spacetime with higher dimensions have attracted a lot of interest in recent years (see e.g., the references cited in \cite{6}). Motivated by the Randall-Sundrum type brane world model \cite{7,8} which was proposed to solve the  hierarchy problem between the Planck and the electroweak scale, anti-de Sitter spacetime has attracted a great deal of attention. The Casimir effect of bosonic fields or massless fermionic fields in anti-de Sitter or Randall-Sundrum spacetime have been considered in \cite{9,10,11,12,13,14,15,16,17,18,19,20,23}. For massive fermionic fields in AdS$_5$ or the Randall-Sundrum spacetime, the Casimir effect has been discussed in \cite{21,22}. In \cite{2}, the Casimir effect of a massive fermionic field in anti-de Sitter spacetime of arbitrary dimensions were computed.  In this letter, we extend the results of \cite{2} by taking into consideration the effect of nonzero temperature. In fact, the Casimir effect of fermionic system at finite temperature has been of considerable interest recently \cite{24,25,26} since such system plays an important role in condensed matter.

The metric of the $(D+1)$-dimensional anti-de Sitter space times AdS$_{D+1}$ is given by
\begin{equation*}
ds^2=g_{\mu\nu}dx^{\mu}dx^{\nu}=e^{-2y/a}\eta_{ik}dx^idx^k-dy^2
\end{equation*}
where $y=x^D$ is the radial coordinate, $i, k = 0,\ldots, D-1$ and $\eta_{ik}=\text{diag}(1,-1,\ldots,-1)$ is the metric tensor of the $D$-dimensional Minkowski spacetime. The parameter $a$ is the AdS curvature radius. In terms of conformal metric $z=ae^{y/a}$,
$$ds^2=\frac{a^2}{z^2}\left(\eta_{jk}dx^jdx^k-dz^2\right).$$The hypersurfaces $z=0$ and $z=\infty$ correspond respectively to the AdS boundary and the horizon.

 The equation of motion of a massive fermionic field with mass $m$  is the Dirac equation:
\begin{equation*}
i\gamma^{\mu}\nabla_{\mu}\psi-m\psi=0,
\end{equation*}
where
$
\nabla_{\mu}=\pa_{\mu}+\Gamma_{\mu},
$ and $\Gamma_{\mu}$ is the spin connection.

We are interested in studying the finite temperature Casimir effect when two boundaries are located at $z=z_1$ and $z=z_2$, with $z_1<z_2$. On the boundaries, we impose the MIT bag boundary conditions:
\begin{equation}\label{eq5_16_2}
(1+i\gamma^{\mu}n_{\mu}^j)\psi\Bigr|_{z=z_j}=0,
\end{equation}where $j=1,2$, and $n_{\mu}^j=(-1)^j\delta_{\mu}^Da/z$ is the  unit  vector normal to the boundary.

Let
$\bar{\gamma}^{\alpha}$ be the Dirac matrices in the $(D+1)$-dimensional Minskowski spacetime. They are matrices of size $N_D\times N_D$, where $N_D=2^{[(D+1)/2]}$, which satisfy $\bar{\gamma}^{\alpha}\bar{\gamma}^{\beta}+\bar{\gamma}^{\beta}\bar{\gamma}^{\alpha}=2\eta^{\alpha\beta}$. In Appendix \ref{a1}, we show that there is a representation of these matrices which assumes the form:
\begin{equation*}
\bar{\gamma}^0=i\begin{pmatrix} 0 & -I\\I & 0\end{pmatrix},\quad\gamma^j=i\begin{pmatrix} 0 & \sigma_j\\\sigma_j & 0\end{pmatrix},\quad j=1,\ldots, D-1,\quad \gamma^D=i\begin{pmatrix} I & 0 \\0 & -I\end{pmatrix}.
\end{equation*}
Here $\sigma_1,\ldots,\sigma_{D-1}$ are $(N_D/2)\times (N_D/2)$ matrices satisfying $\sigma_j\sigma_k+\sigma_k\sigma_j=2\delta_{jk}$. As will be demonstrated below, this representation of Dirac matrices will greatly simplify the analysis of eigenmodes of the Dirac equation, compare to the one used in \cite{2}.

The Dirac matrices $\gamma^{\mu}$ in AdS$_{D+1}$ can be expressed in terms of  $\bar{\gamma}^{\alpha}$ by
$\gamma^{\mu}=e^{\mu}_{\alpha}\bar{\gamma}^{\alpha},$ where $e_{\alpha}^{\mu}$ are the tetrad fields satisfying $e^{\mu}_{\alpha}e^{\nu}_{\beta}\eta^{\alpha\beta}=g^{\mu\nu}$, which can be taken to be
$$e^{\mu}_{\alpha}=\delta^{\mu}_{\alpha}\frac{z}{a}.$$ The spin connection $\Gamma_{\mu}$ is then given by
\begin{gather*}
\Gamma_j=\frac{\eta_{jj}}{2z}\bar{\gamma}^D\bar{\gamma}^j,\quad j=0,1,\ldots,D-1;\hspace{1cm}
\Gamma_D=0.
\end{gather*}
The fermionic field has positive and negative energy modes. For positive modes, write the field as
\begin{equation}
\psi^{(+)}=\begin{pmatrix}\psi_+^{(+)}(z)\\\psi_-^{(+)}(z)\end{pmatrix} e^{i\mathbf{k} \mathbf{x}-i\omega t},
\end{equation}
where
$\mathbf{k}=(k_1,\ldots,k_{D-1})$, $\mathbf{x}=(x^1,\ldots,x^{D-1})$ and $\mathbf{k}\mathbf{x}=k_1x^1+\ldots+k_{D-1}x^{D-1}$. The Dirac equation becomes
\begin{equation}\label{eq5_16_1}
\left[ \left(\pa_z-\frac{D}{2z}\right)\pm \frac{ma}{z}\right]\psi_{\pm}^{(+)}=i\left(-\omega\mp\sum_{j=1}^{D-1}k_j\sigma_j\right)\psi_{\mp}^{(+)}.
\end{equation}
From this, we obtain
\begin{equation*}
\left(\frac{\pa^2}{\pa z^2}-\frac{D}{z}\frac{\pa}{\pa z}+\lambda^2+\frac{D^2+2D}{4z^2}-\frac{m^2a^2}{z^2}\mp \frac{ma}{z^2} \right)\psi_{\pm}^{(+)}=0,
\end{equation*}
where $\lambda^2=\omega^2-k_{\perp}^2$, $k_{\perp}=\sqrt{\sum_{j=1}^{D-1}k_j^2}$. The solution of this equation is given by
\begin{equation}\label{eq5_23_1}
\psi_{\pm}^{(+)}(z)=z^{\frac{D+1}{2}}\left(A_{\pm}^{(+)}J_{ma\pm\frac{1}{2}}(\lambda z)+B_{\pm}^{(+)}H^{(1)}_{ma\pm\frac{1}{2}}(\lambda z)\right),
\end{equation}where $J_{\nu}(z)$ and $H^{(1)}_{\nu}(z)$ are the Bessel function and the Hankel function of the first kind.  Substitute \eqref{eq5_23_1} into
 \eqref{eq5_16_1}, we find that
 \begin{equation}\label{eq5_16_3}\begin{split}
 \lambda A_+^{(+)}=&i\left(-\omega-\sum_{j=1}^{D-1}k_j\sigma_j\right)A_-^{(+)},\\
 \lambda B_+^{(+)}=&i\left(-\omega-\sum_{j=1}^{D-1}k_j\sigma_j\right)B_-^{(+)}.
 \end{split}
 \end{equation}
 Now consider the boundary conditions \eqref{eq5_16_2}, which give
$
\psi_+^{(+)}(z_1)=0$ and $ \psi_-^{(+)}(z_2)=0.
$
These imply that
\begin{equation}\label{eq5_23_2}\begin{split}
B_+^{(+)}=&-\frac{J_{ma+\frac{1}{2}}(\lambda z_1)}{H^{(1)}_{ma+\frac{1}{2}}(\lambda z_1)}A_+^{(+)},\hspace{1cm}
B_-^{(+)}=-\frac{J_{ma-\frac{1}{2}}(\lambda z_2)}{H^{(1)}_{ma-\frac{1}{2}}(\lambda z_2)}A_-^{(+)}.
\end{split}
\end{equation}Multiplying $i\left(-\omega-\sum_{j=1}^{D-1}k_j\sigma_j\right)$ to both sides of the second equation of eq. \eqref{eq5_23_2} and using  eq. \eqref{eq5_16_3}, we have
\begin{equation*}
B_+^{(+)}= -\frac{J_{ma-\frac{1}{2}}(\lambda z_2)}{H^{(1)}_{ma-\frac{1}{2}}(\lambda z_2)}A_+^{(+)}.
\end{equation*}Compare to the first equation of \eqref{eq5_23_2}, we find that for the field $\psi^{(+)}$ to be nontrivial,
\begin{equation}\label{eq5_16_4}
H^{(1)}_{ma+\frac{1}{2}}(\lambda z_1)J_{ma-\frac{1}{2}}(\lambda z_2)-J_{ma+\frac{1}{2}}(\lambda z_1)H^{(1)}_{ma-\frac{1}{2}}(\lambda z_2)=0.
\end{equation}This is the dispersion relation for the eigenfrequency $\omega=\sqrt{\lambda^2+k_{\perp}^2}$. From the analysis above, we also see that there are $N_D/2$ degree of freedom in the vector $A_+^{(+)}$. For the negative energy mode $\psi^{(-)}$ which can be written as
\begin{equation}
\psi^{(-)}=\begin{pmatrix}\psi_+^{(-)}(z)\\\psi_-^{(-)}(z)\end{pmatrix} e^{i\mathbf{k} \mathbf{x}+i\omega t},
\end{equation}
 similar analysis gives rise to the same dispersion relation, and there are another $N_D/2$ degree of freedom.

The finite temperature
Casimir free energy of a fermionic field is  given by
\begin{equation}
E_{\text{Cas}}  = \frac{T}{2}\left(\zeta'_T(0) + [\log \mu^2]\zeta_T(0)\right),
\end{equation}
where $\mu$ is a normalization constant with the dimension of mass, and $\zeta_T(s)$ is the thermal zeta function:
\begin{equation*}
\zeta_T(s)=\sum_{\omega\;\text{eigenfrequencies}}\sum_{l=-\infty}^{\infty}\left(\omega^2+\xi_l^2\right)^{-s}.
\end{equation*}Here $\displaystyle\xi_l= 2\pi\left(l+\frac{1}{2}\right) T$  are the Matsubara frequencies.

We are interested in studying the finite temperature Casimir  stress acting on the two boundaries at $z=z_1$ and $z=z_2$. We need to consider the Casimir energy in the three regions $z<z_1$, $z_1<z<z_2$ and $z>z_2$. Denote the corresponding thermal zeta function as $\zeta_T^{L}(s)$, $\zeta_T^{M}(s)$ and $\zeta_T^{R}(s)$ respectively.

As discussed in the previous section, for the region $z_1<z<z_2$, the eigenfrequencies $\omega$ are solutions of the dispersion relation \eqref{eq5_16_4} with multiplicity $N_D$.  Assume that $-L/2\leq x_j\leq L/2$ for $j=1,\ldots, D-1$. Then as $L\gg (z_2-z_1)$,
\begin{equation*}\begin{split}
\zeta_T^M(s)=&\frac{N_DL^{D-1}}{(2\pi)^{D-1}}\sum_{l=-\infty}^{\infty}\int_{-\infty}^{\infty}dk_1\ldots\int_{-\infty}^{\infty}dk_{D-1}\oint_{C}\frac{du}{2\pi i} \left(u^2+k_{\perp}^2+\xi_l^2\right)^{-s}\\&\hspace{4cm}\times\frac{d}{du}\ln\left(uH^{(1)}_{ma+\frac{1}{2}}(u z_1)J_{ma-\frac{1}{2}}(u z_2)-uJ_{ma+\frac{1}{2}}(u z_1)H^{(1)}_{ma-\frac{1}{2}}(u z_2)\right).\end{split}
\end{equation*}Here $C$  is a closed, counterclockwise contour on the complex plane enclosing all the zeros of eq. \eqref{eq5_16_4}. The contour integral with respect to $u$ is the same as the one that appears in studying Casimir effect of concentric spherical shells (see e.g. \cite{3,4}). Via standard contour integration technique, we find that
\begin{equation*}\begin{split}
\zeta_T^M(s)=&\frac{N_DL^{D-1}}{(2\pi)^{D-1}}\frac{\sin\pi s}{\pi}\sum_{l=-\infty}^{\infty}\int_{-\infty}^{\infty}dk_1\ldots\int_{-\infty}^{\infty}dk_{D-1}\int_{\sqrt{k_{\perp}^2+\xi_l^2}}^{\infty}d\xi \left(\xi^2-k_{\perp}^2-\xi_l^2\right)^{-s}\\&\hspace{4cm}\frac{d}{d\xi}\ln\left(\xi K_{ma+\frac{1}{2}}(\xi z_1)I_{ma-\frac{1}{2}}(\xi z_2)+\xi I_{ma+\frac{1}{2}}(\xi z_1)K_{ma-\frac{1}{2}}(\xi z_2)\right).\end{split}
\end{equation*}Here $I_{\nu}(z)$ and $K_{\nu}(z)$ are modified Bessel functions of first and second kind. Now integrating over $k_1,\ldots, k_{D-1}$,  we have
\begin{equation}\label{eq5_22_5}\begin{split}
\zeta_T^M(s)=&\frac{N_DL^{D-1}}{2^{D-2}\pi^{\frac{D-1}{2}}}\frac{1}{\Gamma(s)\Gamma\left(\frac{D+1}{2}-s\right)}\sum_{l=0}^{\infty} \int_{\xi_l}^{\infty}d\xi  (\xi^2-\xi_l^2)^{\frac{D-1}{2}-s}\\& \times\left\{\frac{d}{d\xi}\ln\left(\xi^{ma+\frac{1}{2}} K_{ma+\frac{1}{2}}(\xi z_1)\right)
+\frac{d}{d\xi} \ln\left(\xi^{-ma+\frac{1}{2}} I_{ma-\frac{1}{2}}(\xi z_2)\right)+\frac{d}{d\xi}\ln\left(1+\frac{ I_{ma+\frac{1}{2}}(\xi z_1)K_{ma-\frac{1}{2}}(\xi z_2)}{K_{ma+\frac{1}{2}}(\xi z_1)I_{ma-\frac{1}{2}}(\xi z_2)}\right)\right\}.\end{split}
\end{equation}
We see that
$\zeta_T^M(s)$ can be decomposed into a sum of three terms:
$$\zeta_T^M(s)=\zeta_{T}^{ML}(s)+\zeta_{T}^{MR}(s)+\zeta_T^{\text{int}}(s),$$corresponding respectively to the three terms in the brackets of eq. \eqref{eq5_22_5}.
$\zeta_T^{ML}(s)$ gives rise to the Casimir free energy in the region $z>z_1$ in the absence of the boundary $z=z_2$,  $\zeta_T^{MR}(s)$ gives rise to the Casimir free energy in the region $z<z_2$ in the absence of the boundary $z=z_1$, and $\zeta_T^{\text{int}}(s)$   gives rise to the interacting free energy between the two boundaries. It follows that the thermal zeta function $\zeta_T^L(s)$ of the region $z<z_1$ is obtained from $\zeta_T^{MR}(s)$ by replacing $z_2$ with $z_1$, and the thermal zeta function $\zeta_T^R(s)$ of the region $z>z_2$ is obtained from $\zeta_T^{ML}(s)$ by replacing $z_1$ with $z_2$. Let
\begin{align*}
\zeta_T^1(s)=&\zeta_T^{L}(s)+\zeta_T^{ML}(s),\hspace{1cm}
\zeta_T^2(s)=\zeta_T^{R}(s)+\zeta_T^{MR}(s).
\end{align*}
Then
\begin{equation}\label{eq5_21_1}\zeta_T^j(s)=\frac{N_DL^{D-1}}{2^{D-2}\pi^{\frac{D-1}{2}}}
\frac{1}{\Gamma(s)\Gamma\left(\frac{D+1}{2}-s\right)}\sum_{l=0}^{\infty} \int_{\xi_l}^{\infty}d\xi  (\xi^2-\xi_l^2)^{\frac{D-1}{2}-s}\frac{d}{d\xi} \ln\left( \xi I_{ma-\frac{1}{2}}(\xi z_j)K_{ma+\frac{1}{2}}(\xi z_j)\right). \end{equation}It gives rise to the Casimir free energy of the single boundary at $z=z_j$.
The total Casimir free energies in the three regions can be written as a sum of three terms:
$$E_{\text{Cas}}=E_{\text{Cas}}^1+E_{\text{Cas}}^2+E_{\text{Cas}}^{\text{int}},$$ which correspond respectively to the thermal zeta functions $\zeta_T^1(s), \zeta_T^2(s)$ and $\zeta_T^{\text{int}}(s)$.

In the following, we only consider the Casimir interaction. It is straightforward to find that $\zeta_T^{\text{int}}(0)=0$, and hence, the Casimir free interaction energy between the two boundaries is given by
\begin{equation*}
\begin{split}
E_{\text{Cas}}^{\text{int}}  =& \frac{T}{2} \zeta_T^{\prime\text{int}}(0)
= -
\frac{N_DL^{D-1}T}{2^{D-2}\pi^{\frac{D-1}{2}}}\frac{1}{ \Gamma\left(\frac{D-1}{2}\right)}\sum_{l=0}^{\infty} \int_{\xi_l}^{\infty}d\xi \,\xi (\xi^2-\xi_l^2)^{\frac{D-3}{2}} \ln\left(1+\frac{ I_{ma+\frac{1}{2}}(\xi z_1)K_{ma-\frac{1}{2}}(\xi z_2)}{K_{ma+\frac{1}{2}}(\xi z_1)I_{ma-\frac{1}{2}}(\xi z_2)}\right).
\end{split}
\end{equation*}Since the modified Bessel functions $I_{\nu}(z)$ and $K_{\nu}(z)$ are always nonnegative for $\nu>-1$, it is obvious that the Casimir free interaction energy between the two boundaries $z=z_1$ and $z=z_2$ is always negative.

The area of the hypersurface at $z=z_j$ with $-L/2\leq x_k\leq L/2$, $k=1,\ldots,D-1$, is $A_j=(La/z_j)^{D-1}$. Hence, the Casimir pressure (or Casimir force density) acting on the boundary $z=z_j$ induced by the interaction free energy is given by
\begin{equation*}
\begin{split}
P_{\text{Cas}}^{\text{int},j}=&-\frac{1}{A_j}\left(\frac{z_j}{a}\right)^{2}\frac{\pa}{\pa z_j}E_{\text{Cas}}^{\text{int}} \\
=&-\frac{1}{L^{D-1}}\left(\frac{z_j}{a}\right)^{D+1}\frac{\pa}{\pa z_j}E_{\text{Cas}}^{\text{int}},\end{split}\end{equation*}i.e.,
\begin{equation}\label{eq5_22_4}
\begin{split}
P_{\text{Cas}}^{\text{int},1}=&\frac{N_D(z_1/a)^{D+1}T}{2^{D-2}\pi^{\frac{D-1}{2}}}\frac{1}{ \Gamma\left(\frac{D-1}{2}\right)}\sum_{l=0}^{\infty} \frac{1}{z_1}\int_{\xi_l}^{\infty}d\xi \,\xi (\xi^2-\xi_l^2)^{\frac{D-3}{2}} \\&\hspace{3cm}\times \frac{ K_{ma-\frac{1}{2}}(\xi z_2)}{K_{ma+\frac{1}{2}}(\xi z_1)}\frac{1}{I_{ma+\frac{1}{2}}(\xi z_1)K_{ma-\frac{1}{2}}(\xi z_2)+K_{ma+\frac{1}{2}}(\xi z_1)I_{ma-\frac{1}{2}}(\xi z_2)},\\
P_{\text{Cas}}^{\text{int},2}=&-\frac{N_D(z_2/a)^{D+1}T}{2^{D-2}\pi^{\frac{D-1}{2}}}\frac{1}{ \Gamma\left(\frac{D-1}{2}\right)}\sum_{l=0}^{\infty} \frac{1}{z_2}\int_{\xi_l}^{\infty}d\xi \,\xi (\xi^2-\xi_l^2)^{\frac{D-3}{2}}  \\&\hspace{3cm}\times\frac{ I_{ma+\frac{1}{2}}(\xi z_1)}{I_{ma-\frac{1}{2}}(\xi z_2)}\frac{1}{I_{ma+\frac{1}{2}}(\xi z_1)K_{ma-\frac{1}{2}}(\xi z_2)+K_{ma+\frac{1}{2}}(\xi z_1)I_{ma-\frac{1}{2}}(\xi z_2)}.
\end{split}\end{equation}As have been noticed in \cite{2}, the Casimir pressure on the two boundaries are in general different due to the nontrivial curvature. Nevertheless, one can immediately observe from \eqref{eq5_22_4} that $P_{\text{Cas}}^{\text{int},1}$ is always positive, and $P_{\text{Cas}}^{\text{int},2}$ is always negative. In other words, the pressure induced by the interaction energy is attractive, i.e., tends to bring the two boundaries closer to each other.

In the high temperature limit, i.e., $T\Delta z=T(z_2-z_1)\gg 1$, since the integrands in \eqref{eq5_22_4} behaves like $e^{-2(z_2-z_1)\xi}$ for large $\xi$, and $\xi_l=2\pi(l+1/2)  T\rightarrow \infty$ as $T\rightarrow\infty$, we find that the Casimir interaction decays exponentially in the high temperature limit. This is in sharp contrast with finite temperature Casimir effect of bosonic fields, where the high temperature limit of the interaction is linear in temperature, contributed from the terms with zero Matsubara frequency. In the fermionic case, the Matsubara frequencies are all nonzero.

In the low temperature limit $T\Delta z=T(z_2-z_1)\ll 1$,  the leading term is the zero temperature term obtained by replacing the summation over $\xi_l$ in \eqref{eq5_22_4} by an integration, i.e.,
\begin{equation*}
\sum_{l=0}^{\infty} \int_{\xi_l}^{\infty}d\xi \,\xi (\xi^2-\xi_l^2)^{\frac{D-3}{2}}\longrightarrow \frac{1}{2\pi T}\int_0^{\infty} du \int_{u}^{\infty}d\xi \,\xi (\xi^2-u^2)^{\frac{D-3}{2}}=\frac{1}{4\sqrt{\pi}T}\frac{\Gamma\left(\frac{D-1}{2}\right)}{\Gamma\left(\frac{D}{2}\right)}\int_0^{\infty} d\xi,
\end{equation*}and we obtain the same result as in \cite{2}.

\begin{figure}[h]
\epsfxsize=0.49\linewidth \epsffile{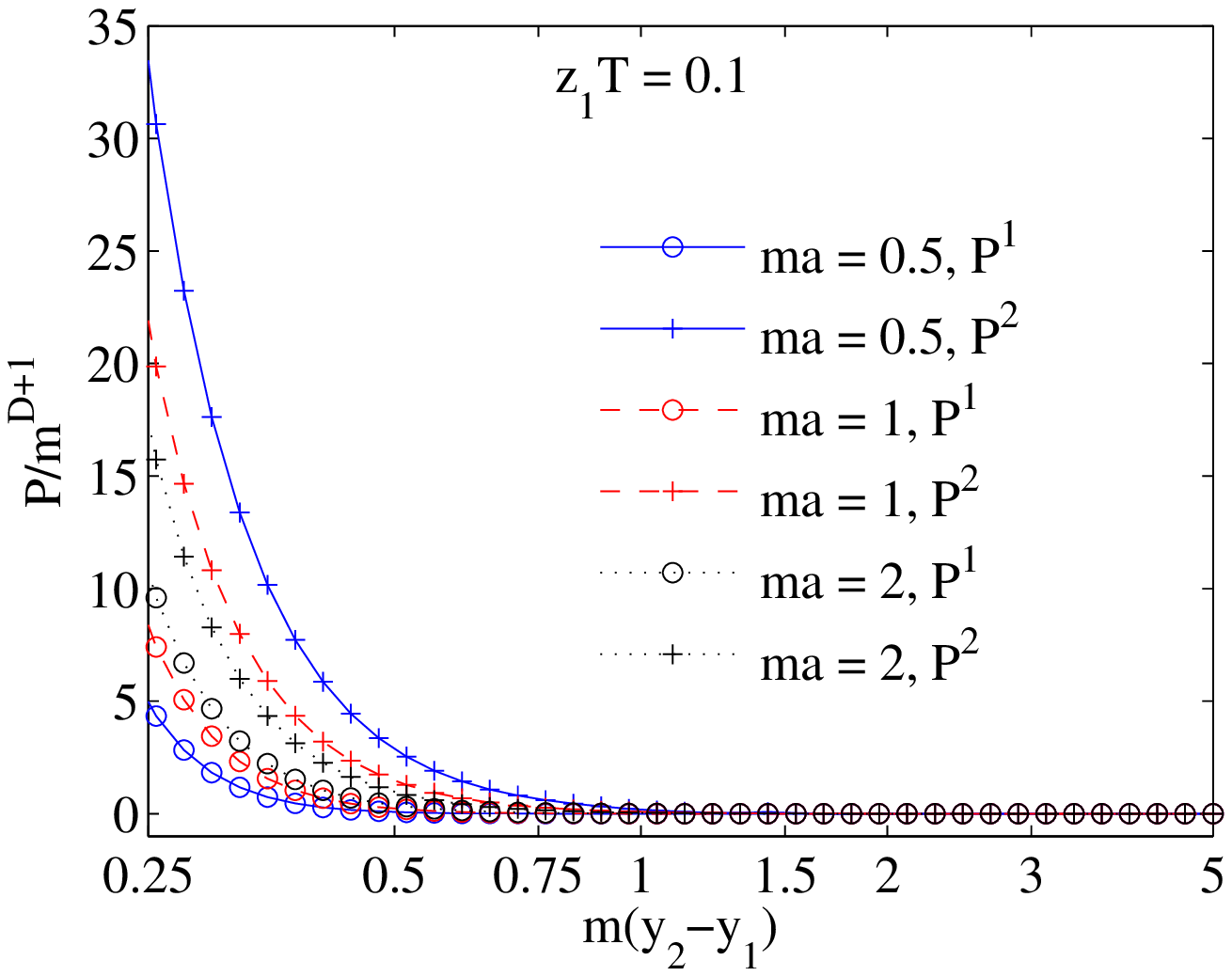}  \epsfxsize=0.49\linewidth \epsffile{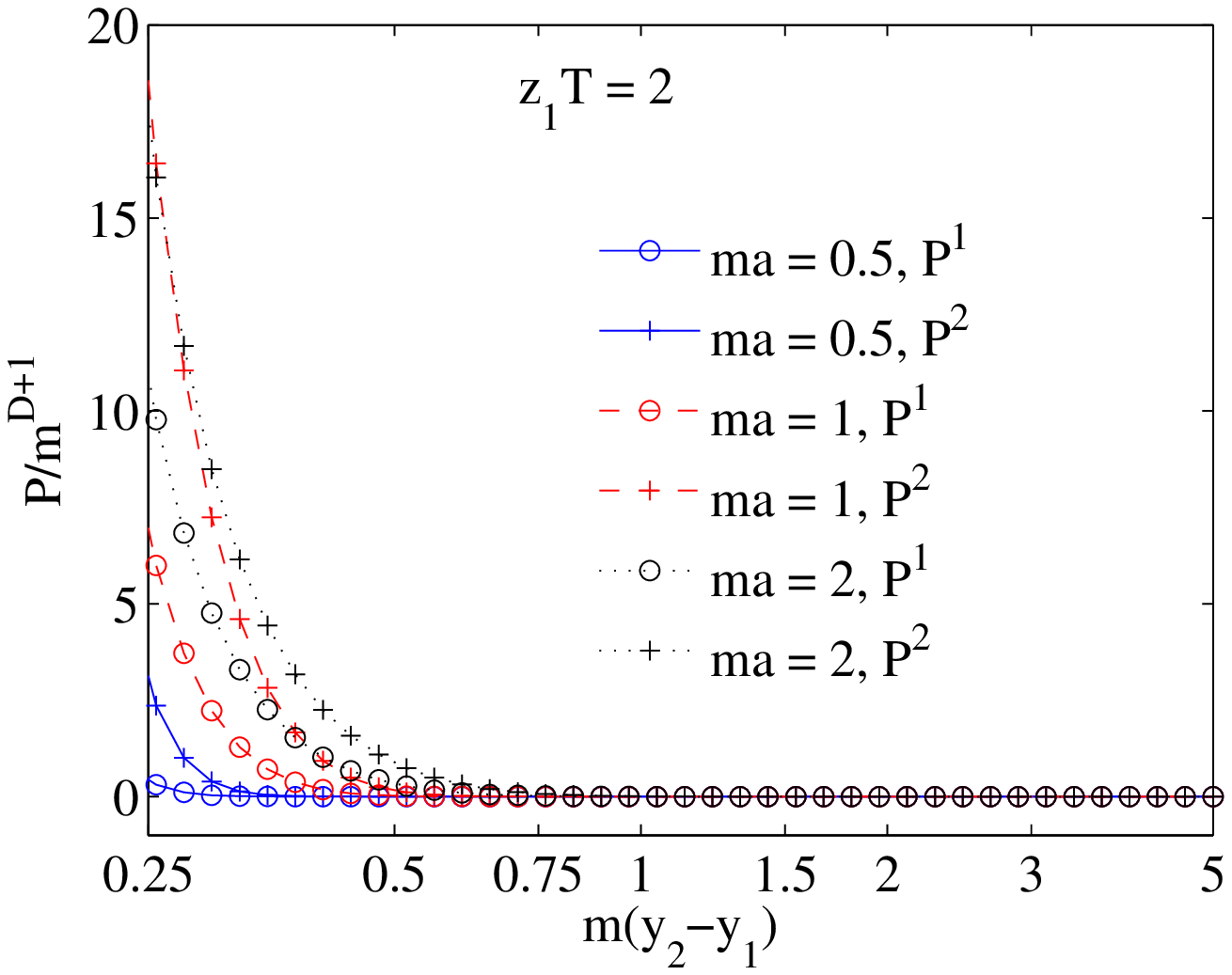}\caption{\label{f1} The normalized pressure $P_{\text{Cas}}^{\text{int},j}/m^{D+1}$ as a function of $m(y_2-y_1)$ for $z_1T=0.1$ and $z_1T=2$. }\end{figure}

\begin{figure}[h]
\epsfxsize=0.49\linewidth \epsffile{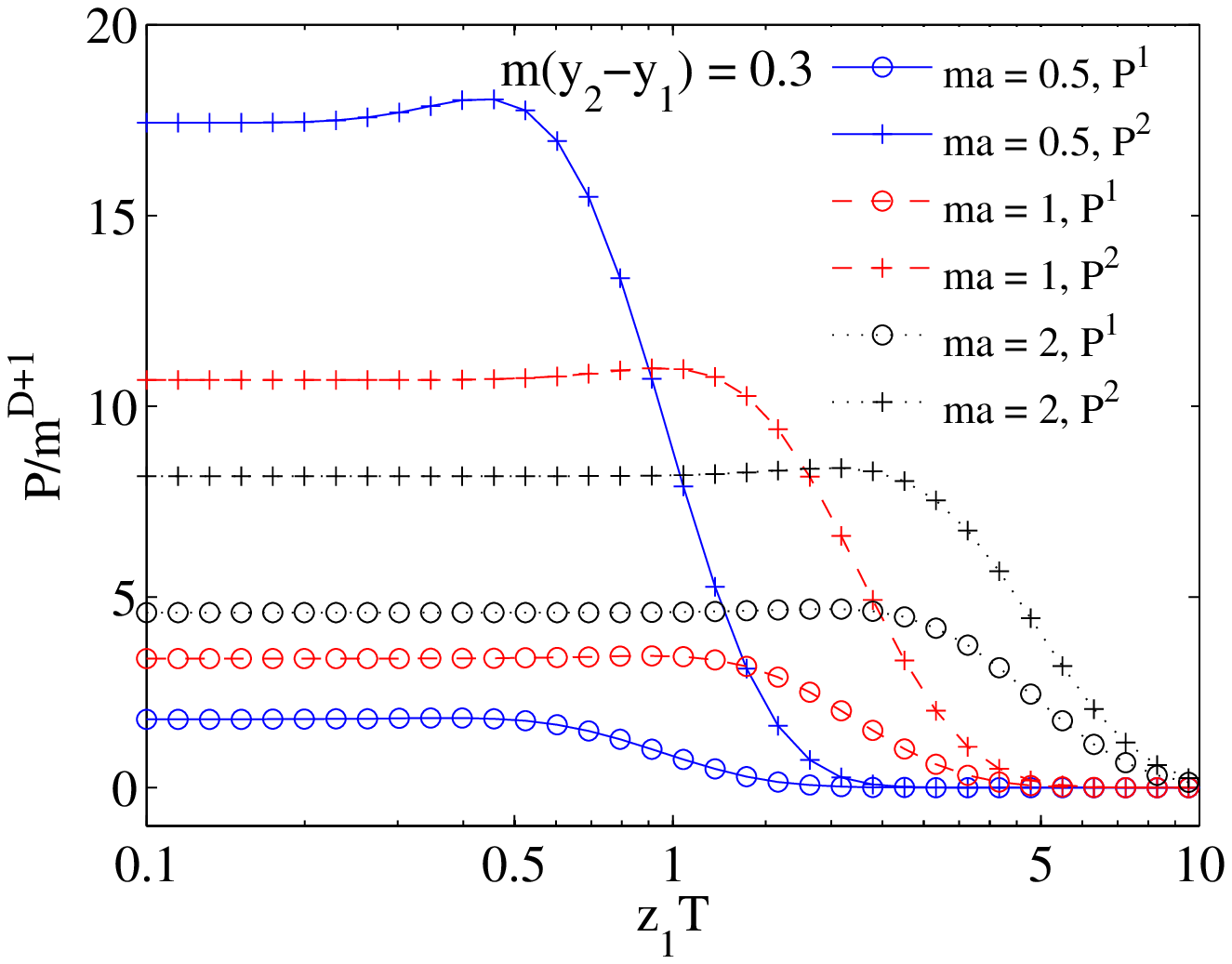}  \epsfxsize=0.49\linewidth \epsffile{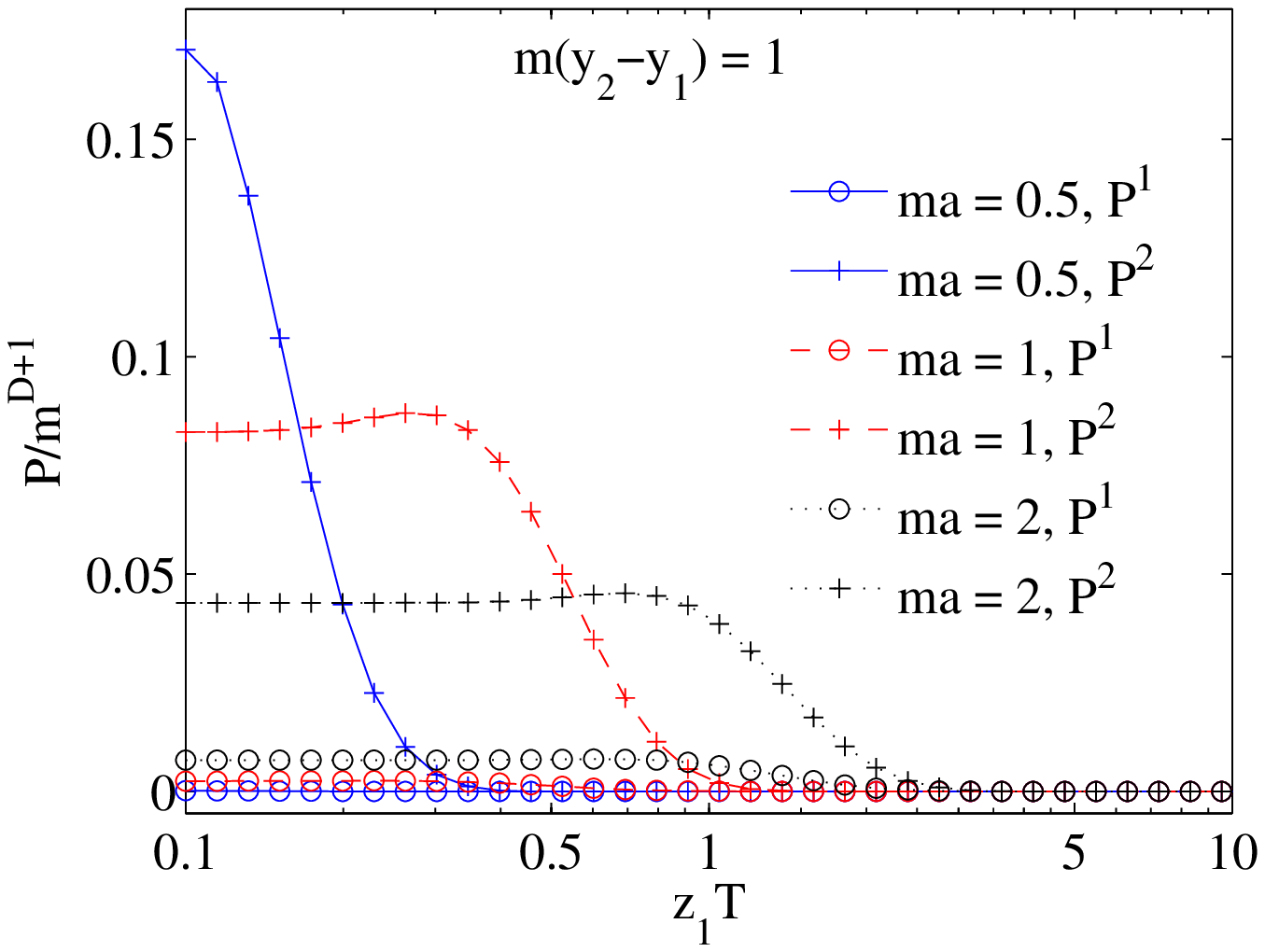}\caption{\label{f2} The normalized pressure $P_{\text{Cas}}^{\text{int},j}/m^{D+1}$ as a function of $z_1T$ for $m(y_2-y_1)=0.3$ and $m(y_2-y_1)=1$. }\end{figure}

In the Minkowskian limit $ma\gg 1$ with $m$ fixed, $r=z_1/z_2\approx 1-(y_2-y_1)/a$ and $\nu/z_2\approx m$. Making a change of variables $\xi\mapsto \nu\xi/z_2$, Debye uniform asymptotic expansions of modified Bessel functions \cite{5} give
\begin{align*}
E_{\text{Cas}}^{\text{int}}
= &-
\frac{N_DL^{D-1}T}{2^{D-2}\pi^{\frac{D-1}{2}}}\frac{1}{ \Gamma\left(\frac{D-1}{2}\right)}\sum_{l=0}^{\infty}\!'m^{D-1}\int_{ \xi_l/m}^{\infty}d\xi \,\xi (\xi^2-( \xi_l/m)^2)^{\frac{D-3}{2}} \ln\left(1+\frac{\sqrt{\xi^2+1}-1}{\sqrt{\xi^2+1}+1}\exp\left(-2m(y_2-y_1)\sqrt{\xi^2+1}\right)\right).
\end{align*}Making another  change of variables $$m\sqrt{\xi^2+1}=x,$$ we find that
\begin{equation*}
E_{\text{Cas}}^{\text{int}}
\approx -
\frac{N_DL^{D-1}T}{2^{D-2}\pi^{\frac{D-1}{2}}}\frac{1}{ \Gamma\left(\frac{D-1}{2}\right)}\sum_{l=0}^{\infty}\!' \int_{ \sqrt{\xi_l^2+m^2}}^{\infty}dx \,x(x^2-\xi_l^2-m^2)^{\frac{D-3}{2}} \ln\left(1+\frac{x-m}{x+m}\exp\left(-2x\Delta y  \right)\right),
\end{equation*}where $\Delta y=y_2-y_1$. This agrees with the result obtained in \cite{6} for the Casimir free energy between two parallel plates separated by a distance $\Delta y$ in $(D+1)$ dimensional Minkowski spacetime.

For $D=3$, the graphs of the Casimir pressure $P_{\text{Cas}}^{\text{int},j}$ are shown in Fig. \ref{f1} and Fig. \ref{f2}.  In Fig. \ref{f1}, we plot the normalized pressure $P^{\text{int}}/m^{D+1}$ as a function of $m(y_2-y_1)$ for $ma=0.5, 1, 2$ and $z_1T=0.1,2$. As are shown by the graphs, the Casimir pressure decreases when $m(y_2-y_1)$ increases. We also observe that the magnitude of the Casimir pressure at $z_1T=2$ is smaller than at $z_1T=0.1$, and the reduction in magnitude is more significant for smaller $ma$. The dependence of the normalized Casimir pressure on $z_1T$ is shown in Fig. \ref{f2} for $m(y_2-y_1)=0.3$ and $m(y_2-y_1)=1$. As discussed above, we find that the Casimir pressure decays  to 0 at large $z_1T$. However, it does not always decrease monotonically.

We have investigated the Casimir interaction at any finite temperature between two parallel boundaries in anti-de Sitter spacetime of any dimension which is induced by the vacuum fluctuations of a massive fermionic field subject to MIT bag boundary conditions on the two boundaries. A   representation of Dirac matrices is introduced which greatly simplify the analysis of eigenmodes of the field in these type of problems. The Casimir interaction energy and Casimir pressure is computed using standard contour integration technique and zeta regularization. It was established that at any temperature and for any mass parameter, the Casimir pressure between the two boundaries always tends to bring the two boundaries closer to each other. In the zero temperature limit, we recover the result of \cite{2}. In the high temperature limit, we find that the Casimir interaction decays exponentially. This is in drastic contrast to Casimir effect of bosonic fields whose high temperature leading term is linear in temperature. The reason for such contrast is because there is no Matsubara frequency that is equal to zero in the fermionic case.

\begin{acknowledgments}\noindent
  This work is supported by the Ministry of Higher Education of Malaysia  under   FRGS grant FRGS/1/2013/ST02/UNIM/02/2.
  \end{acknowledgments}
\appendix
\section{Dirac matrices}\label{a1}
In this section, we construct explicitly Dirac matrices $\bar{\gamma}^0,\ldots,\bar{\gamma}^D$ in $(D+1)$-dimensional Minkowski spacetime, which are matrices of size $N_D\times N_D$, with $N_D=2^{[(D+1)/2]} $,   having the following specific representation:
\begin{equation*}
\bar{\gamma}^0=i\begin{pmatrix} 0 & -I\\I & 0\end{pmatrix},\quad\gamma^j=i\begin{pmatrix} 0 & \sigma_j\\\sigma_j & 0\end{pmatrix},\quad j=1,\ldots, D-1,\quad \gamma^D=i\begin{pmatrix} I & 0 \\0 & -I\end{pmatrix}.
\end{equation*}
Here $\sigma_1,\ldots,\sigma_{D-1}$ are $(N_D/2)\times (N_D/2)$ matrices satisfying $\sigma_j\sigma_k+\sigma_k\sigma_j=2\delta_{jk}$. It follows that for $\alpha,\beta=0,1,\ldots, D$,
$$\bar{\gamma}^{\alpha}\bar{\gamma}^{\beta}+\bar{\gamma}^{\beta}\bar{\gamma}^{\alpha}=2\eta^{\alpha\beta}.$$

When $D=1$, let
\begin{align*}
\bar{\gamma}^0= i\begin{pmatrix} 0 & -1\\
1 & 0\end{pmatrix},\hspace{1cm}\bar{\gamma}^{1}= i\begin{pmatrix} 1 & 0\\0 & -1\end{pmatrix}.
\end{align*}

When $D=2$, let
\begin{align*}
\bar{\gamma}^0= i\begin{pmatrix} 0 & -1\\
1 & 0\end{pmatrix},\hspace{1cm}\bar{\gamma}^{1}=i\begin{pmatrix} 0 & 1\\1 & 0\end{pmatrix},\hspace{1cm}\bar{\gamma}^{2}=i \begin{pmatrix} 1 & 0\\0& -1\end{pmatrix}.
\end{align*}

When $D\geq 3$, take the $(D-1)$ Dirac matrices $\tilde{\gamma}^0,\ldots,\tilde{\gamma}^{D-2}$ in $(D-1)$ dimensional Minkowski spacetime. Then let
\begin{align*}
\bar{\gamma}^0=i\begin{pmatrix} 0 & -I \\ I & 0\end{pmatrix},\quad \bar{\gamma}^1=i\begin{pmatrix}0 & \tilde{\gamma}^0  \\ \tilde{\gamma}^0  & 0\end{pmatrix},\quad
\bar{\gamma}^{j+1}=i\begin{pmatrix}  0 & i\tilde{\gamma}^j \\ i\tilde{\gamma}^j & 0 \end{pmatrix}, j=1,\ldots, D-2,
\end{align*}
and
$$\bar{\gamma}^{D}=i\begin{pmatrix}   I & 0\\0 & -I  \end{pmatrix}.$$It is easy to verify by induction that these matrices have the required properties.

\end{document}